\documentclass{PoS}

\usepackage[latin9]{inputenc}
\usepackage{amsmath}
\usepackage{amssymb}
\usepackage{textcomp}
\usepackage{graphicx}
\DeclareGraphicsExtensions{.pdf,.eps}
\usepackage{subfig}
\usepackage{xcolor}

\title{LHC and DIS experimental data \linebreak in the CT18(Z) global QCD analysis}

\ShortTitle{Experimental data in the CT18(Z) analysis}

\author{Tie-Jiun Hou${}^1$, Keping Xie${}^{2,9}$, Jun Gao${}^{3,4}$, Sayipjamal Dulat${}^5$, Marco Guzzi${}^6$, \linebreak T.~J. Hobbs${}^{2,7}$, Joey Huston${}^8$, \speaker{Pavel Nadolsky}\footnote{E-mail: nadolsky@smu.edu}\ ${}^2$, Jon Pumplin${}^8$, Carl Schmidt${}^8$, Ibrahim Sitiwaldi${}^5$, Dan Stump${}^8$, 
Bo-Ting Wang${}^2$, C.-P. Yuan${}^8$\\
	${}^1$Department of Physics, College of Sciences, Northeastern University, Shenyang 110819, China\\
${}^2$Southern Methodist University, Dallas, TX 75275-0175, U.S.A.\\
	${}^3$Shanghai Jiao Tong University, Shanghai 200240, China \\
	${}^4$Center for High Energy Physics, Peking University, Beijing 100871, China \\
	${}^5$Xinjiang University, Urumqi, Xinjiang 830046, China\\
	${}^6$Kennesaw State University, GA 30144 , U.S.A.\\
${}^{7}$Jefferson Lab, EIC Center, Newport News, VA 23606, U.S.A.\\
	${}^8$Michigan State University, East Lansing, MI 48824, U.S.A.
	${}^9$Department of Physics and Astronomy, University of Pittsburgh, Pittsburgh, PA 15260, USA
}

\abstract{We discuss implementation of the LHC experimental data sets
  in the new CT18 global analysis of quantum chromodynamics (QCD) at
  the next-to-next-leading order of the QCD coupling strength. New
  methodological developments in the fitting methodology are
  discussed. Behavior of the CT18 NNLO PDFs for the conventional and
  ``saturation-inspired'' factorization scales in deep-inelastic
  scattering is reviewed. Four new families of (N)NLO CTEQ-TEA PDFs
  are presented: CT18, A, X, and Z.}

\FullConference{XXVII International Workshop on Deep-Inelastic Scattering and Related Subjects - DIS2019\\
		8-12 April, 2019\\
		Torino, Italy}

\dedicated{MSUHEP-19-019, SMU-HEP-19-13}

\begin{document}

%\section{...}
In the companion contribution \cite{Yuan:DIS2019}, we presented the
new CT18 global QCD analysis of parton distribution functions
(PDFs). The CT18 analysis updates widely used CT14 PDF
sets~\cite{Dulat:2015mca} by applying NNLO and NLO global fits to an
expanded set of experimental measurements that include high-luminosity
data from the $ep$ collider HERA and the Large Hadron
Collider. The CT18 experimental data set includes high-statistics measurements from ATLAS, CMS, and LHCb on 
production of inclusive jets, $W/Z$ bosons, and top quark pairs, 
while it retaining crucial {\it legacy} data, such as measurements
from the Tevatron and the HERA Run I and Run II combined data. In this
contribution,
we review implementation of the new data sets in the CT18 global fit 
and the associated physics issues that 
affect the resulting PDFs and a wide class of QCD predictions
based on them. 

By 2018, the LHC collaborations published
about three dozen experimental data sets that can potentially constrain 
the CT PDFs. In light of the unprecedented precision reached in some 
measurements, the latest LHC data must be analyzed using
next-to-next-to-leading order (NNLO) theoretical predictions in
perturbative QCD. The final PDFs depend 
on numerous systematic factors in the experimental data; and the scope
of numerical computations needs to be expanded, too. A systematic examination 
of these effects is essential for trustworthy estimates of PDF uncertainties. 

\textbf{Combined HERA I+II DIS data and an $x$-dependent factorization
  scale.} Even in the 
LHC era, the DIS data from $ep$ collider HERA provides the dominant
constraints on the CT18 PDFs. This dominance can be established using
the \texttt{ePump} and \texttt{PDFSense} statistical techniques reviewed below.
CT18 implements the final (``combined'') data set from DIS at
HERA-I and II \cite{Abramowicz:2015mha} that supersedes the HERA-I
only data set \cite{Aaron:2009aa} used in CT14 \cite{Dulat:2015mca}.
A transitional PDF set, CT14HERA2, was released based on fitting the
final HERA data \cite{Hou:2016nqm}. We found fair overall agreement of the HERA I+II
data with both CT14 and CT14HERA2 PDFs, and that both PDF ensembles
describe equally well the non-HERA data included in our global analysis.
At the same time, we observed some disagreement (``statistical tension'') 
between the $e^{+}p$ and $e^{-}p$ DIS cross sections of the HERA I+II data set. 
We determined that, at the moment, no plausible explanation conclusively explains
the full pattern of these tensions, as they are distributed across
the whole accessible range of Bjorken $x$ and lepton-proton momentum
transfer $Q$ at HERA.

It has been argued that resummation of logarithms $\ln^p(1/x)$ at
$x\ll 1$ improves agreement withe HERA Run I+II data by several tens
of units of $\chi^2$ \cite{Ball:2017otu,Abdolmaleki:2018jln}. In our
analysis, we observe that, by evaluating the DIS cross sections at
NNLO with an $x$-dependent factorization scale, such as a tuned scale
$\mu^2_{F,x} = 0.8^2\ \left(Q^2 + 0.3\mbox{\, GeV}^2/x^{0.3}\right)$,
instead of the conventional choice $\mu^2_F =Q^2$, we achieve a
comparable quality of improvement in the description of the HERA DIS
data set by the {\it fixed-order} NNLO theoretical prediction as the
inclusion of the low-$x$ resummation in
\cite{Ball:2017otu,Abdolmaleki:2018jln}. Namely, the
$\chi^2(\mbox{HERAI+II})$ reduces by $> 50$ units in the kinematical region
$Q > 2\mbox{ GeV}$, $x > 10^{-5}$ of the DIS data included in the CT18
global fit. The parametric form of the $x$-dependent scale
$\mu^2_{F,x}$ is inspired by qualitative saturation arguments (see, e.g.,
\cite{Caola:2009iy}), and the numerical coefficients in $\mu^2_{F,x}$
are chosen to  minimize $\chi^2$ for the HERA DIS data.

\begin{figure}[tb]
	\includegraphics[width=0.59\textwidth]{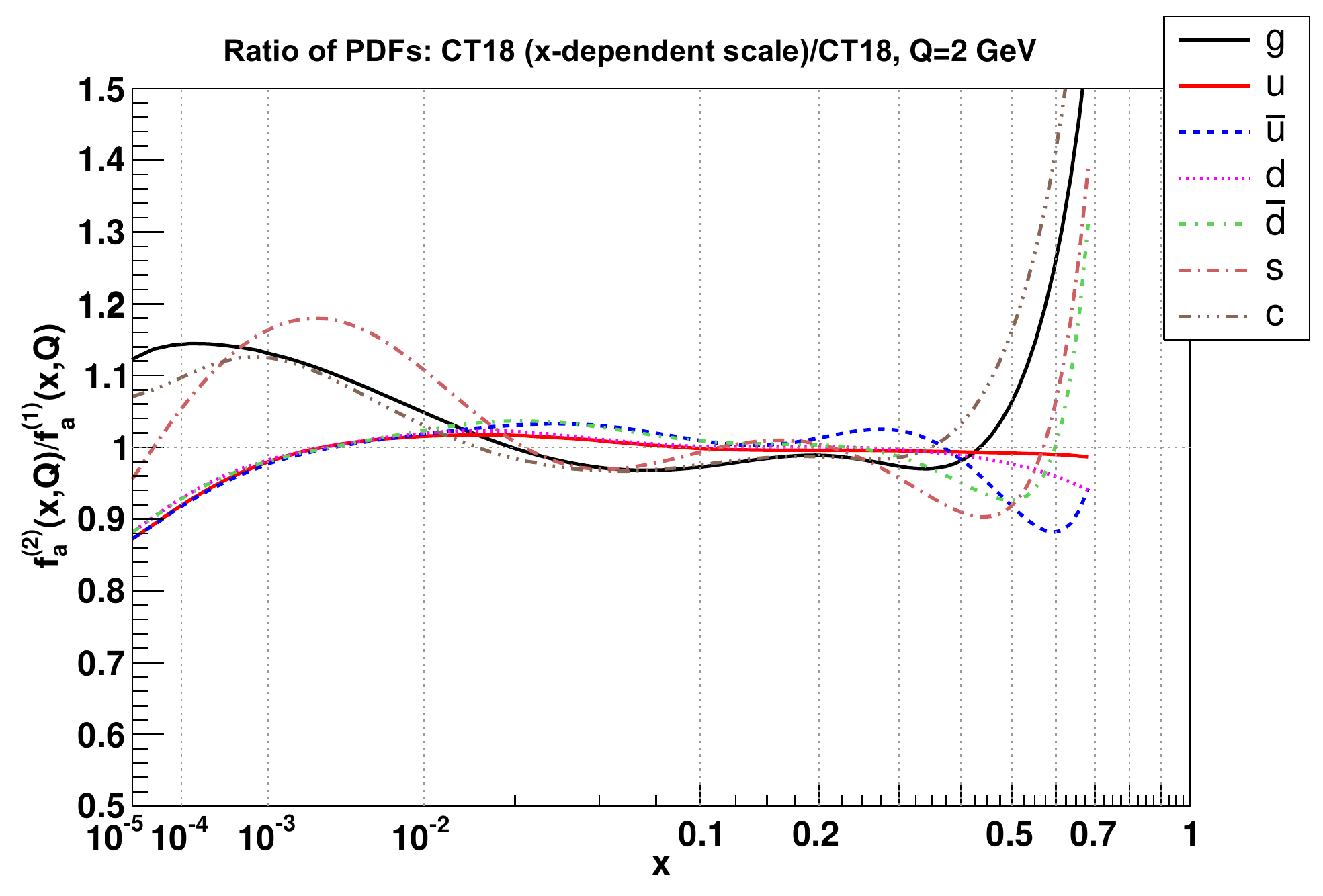}
	\includegraphics[width=0.4\textwidth]{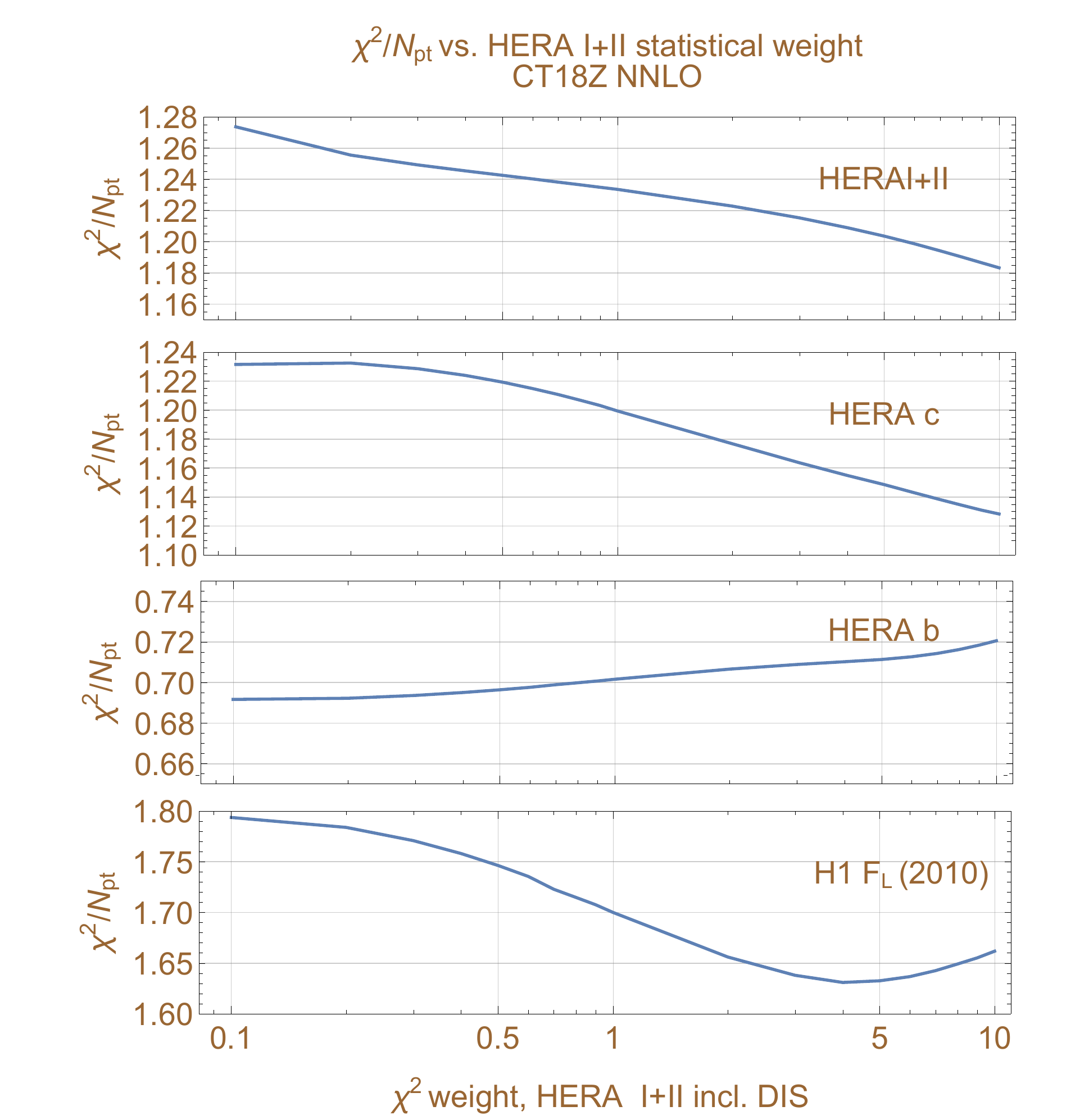}
\caption{Left: The ratios of the candidate CT18 NNLO PDFs obtained with the
  $x$-dependent and standard factorization scales in DIS data
  sets. Right: The $\chi^2/N_{pt}$ values for four HERA data sets in
  the CT18Z fits with the $x$-dependent DIS factorization scale and
  varied statistical weight of the HERA I +II inclusive DIS data set.}
\label{fig:saturation}
\end{figure}

Fig.~\ref{fig:saturation}(left) shows
the changes in the candidate CT18 PDFs obtained by fitting DIS with
the $x$-dependent factorization scale, as compared to the CT18 PDFs
with the nominal scale. With the scale $\mu_{F,x}^2$, we observe
reduced $u$ and $d$ (anti-)quark PDFs and increased gluon and
strangeness PDFs at $x < 10^{-2}$ as compared to the nominal CT18 fit,
with some compensating changes occuring in the same PDFs in the
unconstrained region $x > 0.5$ in order to satisfy the valence and
momentum sum rules. The right Fig.~\ref{fig:saturation} shows the
$\chi^2/N_{pt}$ values (divided by the number $N_{pt}$ of experimental data
points) for four HERA data sets (inclusive NC+CC DIS
\cite{Abramowicz:2015mha},
reduced charm, bottom production cross sections, and H1 longitudinal function
$F_L(x,Q^2)$ \cite{Collaboration:2010ry}) in the fits with the varied
statistical weight $w$ of the HERA I+II inclusive DIS data set \cite{Abramowicz:2015mha}. The
default CT18 fits correspond to $w=1$; with $w=10$, the CT18 fit
increasingly behaves as a HERA-only fit. We see that, with the scale
$\mu^2_{F,x}$ and $w=10$, $\chi^2/N_{pt}$ for the inclusive DIS data set
improves almost to the levels observed in the ``resummed'' HERA-only fits
without intrinsic charm \cite{Ball:2017otu,Abdolmaleki:2018jln}. The
quality of the fit to the charm SIDIS cross section and H1 $F_L$ also improves.

\begin{figure}[tb]
	\includegraphics[width=0.49\textwidth]{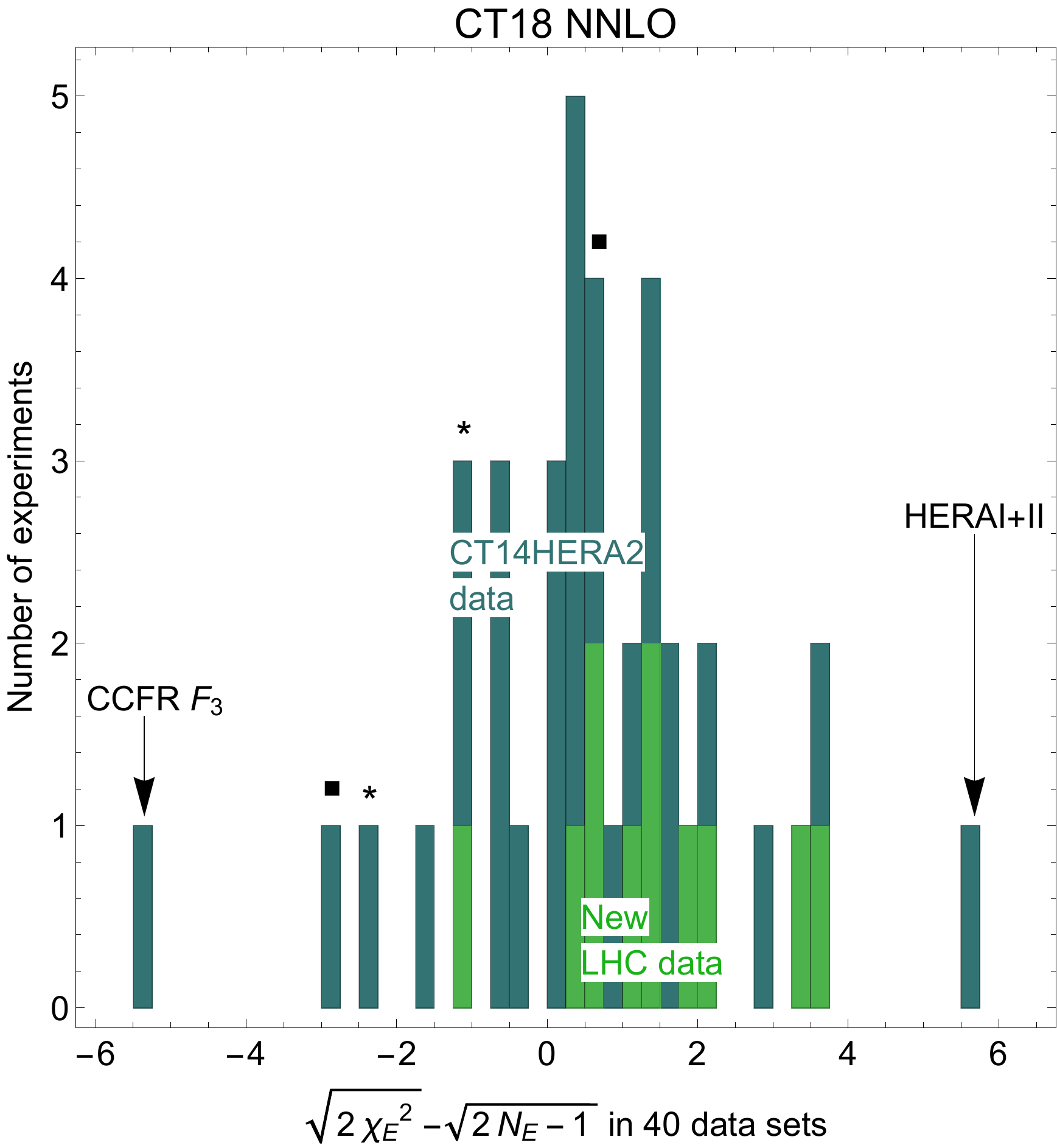}
	\includegraphics[width=0.49\textwidth]{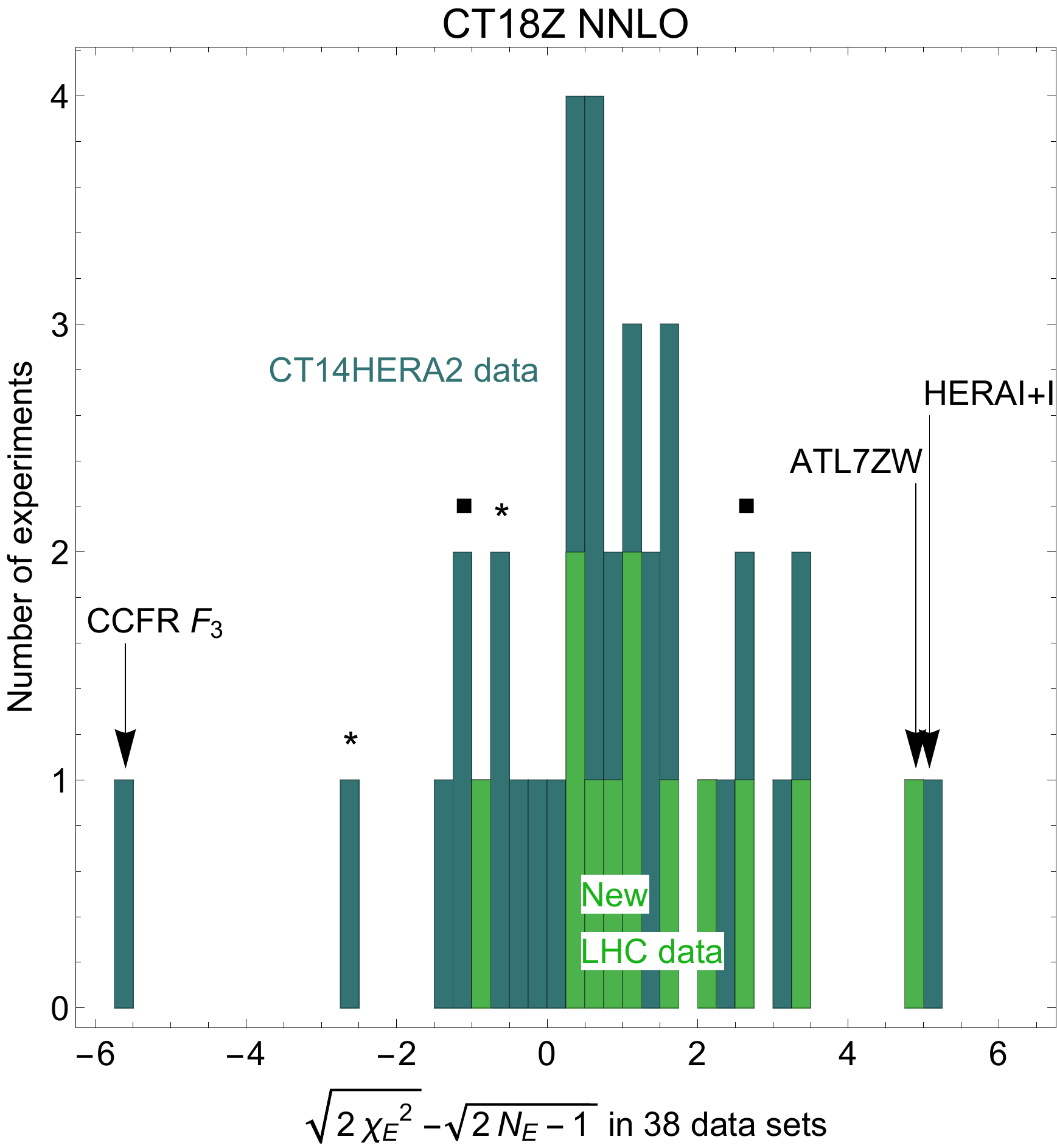}
	\caption{The effective Gaussian variable ($S_n$) distribution of all (a) CT18 data sets, and (b) CT18Z data sets. 
	Two squares and two stars indicate the $S_n$ values for the NuTeV dimuon and CCFR dimuon data, respectively.
\label{fig:sn_ct18_ct18z}}
\end{figure}

\textbf{Selection of new LHC experiments.} When selecting the 
most promising LHC experiments for the CT18 fit, 
we had to address a recurrent challenge, the presence of statistical
tensions among various (sub)sets of the latest experimental data from
HERA, LHC, and the Tevatron. The quickly improving precision of the collider
data reveals previously irrelevant anomalies either in the experiment
or theory. These anomalies are revealed by applying strong goodness-of-fit
tests \cite{Kovarik:2019xvh}. Figure~\ref{fig:sn_ct18_ct18z}
illustrates the degree of tensions using a representation based on the effective
Gaussian variables $S_E\equiv \sqrt{2\chi_E^2}-\sqrt{2 N_E-1}$ \cite{Lai:2010vv}
constructed from the $\chi^2$ values and numbers of data points for
individual data sets $E$. In a high-quality fit, the probability
distribution for $S_E$ must be approximately a standard normal
distribution (with a unit half-width). In CTEQ-TEA and global
fits from either CTEQ or other groups, 
we in fact observe wider $S_E$ distributions,
cf. Fig.~\ref{fig:sn_ct18_ct18z}, with some most comprehensive
and precise data sets (notably, HERA I+II inclusive DIS
\cite{Abramowicz:2015mha} and ATLAS 7 TeV $Z/W$ production
\cite{Aaboud:2016btc}) having $S_E$ values as high as five units or more.  
The question, then, is how to select the
clean and accurate experiments for the global analysis from the list
that grows day-by-day, while maximally preserving the consistency of
the selected experiments. 

For example, there are many LHC experimental data sets \cite{Rojo:2015acz}
that are potentially sensitive to the PDFs, including novel measurements
in production of high-$p_{T}$ $Z$ bosons, $t\bar{t}$ pairs, heavy
quarks, and $W+c$ pairs. Including all such candidate experiments
into the full global fit is impractical: CPU costs grow quickly with
the number of experimental data sets at NNLO. Poorly fitted experiments
would increase, not decrease, the final PDF uncertainty. The generation
of one error PDF set took several days of CPU time in the CT14 fit
to 33 experiments in a single-thread mode. Adding 20-30 additional
experiments with this setup was thus impossible.

\textbf{Advancements in fitting methodology.} The CTEQ-TEA group resolved
these challenges through a multi-prone effort. We developed two programs
for fast preliminary analysis to identify the eligible experimental
data sets for the global fit. The \texttt{PDFSense} program
\cite{Wang:2018heo} was developed at SMU
to predict quantitatively, and before doing the fit, which data sets
will have an impact on the global PDF fit. The \texttt{ePump} program
\cite{Schmidt:2018hvu} developed at MSU applies PDF reweighting to
quickly estimate the impact of data on the PDFs prior to the global
fit. These programs provide helpful guidelines for the selection of the 
most valuable experiments based entirely on the previously published
Hessian error PDFs.

The CTEQ fitting code was parallelized to allow faster turnaround
time (one fit within few hours instead of many days) on high-performance
computing clusters. For as much
relevant LHC data as possible, we computed our own tables for
APPLGrid/fastNLO fast
interfaces \cite{Kluge:2006xs,Carli:2010rw} for NLO cross sections
(to be multiplied by tabulated point-by-point NNLO/NLO corrections)
for various new LHC processes: production of high-$p_{T}$
Z bosons, jets, $t\bar{t}$ pairs. The APPLgrid tables were cross validated
against similar tables from other groups (available in the public
domain) and optimized for speed and accuracy. 

\textbf{The resulting family of new PDFs} consists of four NNLO PDF
ensembles, and the corresponding NLO ones: the default CT18 ensemble and three
alternative ensemble, designated as CT18A, X, and Z. 
Based on the \texttt{PDFSense} and \texttt{ePump}
studies, eleven new LHC data sets have been included in all four PDF fits,
notably, data at 7 and 8 TeV on lepton pair, jet, and $t\bar{t}$
production. Significant effort was spent on understanding the sources
of PDF uncertainties. Theoretical uncertainties associated with the
scale choice were investigated for the affected processes such as
DIS and high-$p_{T}$ $Z$ production. Other considered theoretical
uncertainties were due to the differences among the NNLO/resummation
codes (FEWZ, ResBos, MCFM, NNLOJet++,...) and Monte-Carlo integration.
The important parametrization uncertainty was investigated by repeating
the fits for 90+ trial functional forms of the PDFs. [Our post-CT10
fits parametrize PDFs using Bernstein polynomials, which simplify
trying a wide range of parametrization forms to quantify/eliminate
potential biases.] In addition to the default CT18 PDF ensemble,
the other three sets were obtained under alternative assumptions.
(a) The CT18A and CT18Z analyses include high-luminosity ATLAS 7 TeV $W/Z$
rapidity distributions \cite{Aaboud:2016btc} that show some tension
with DIS experiments and prefer a larger strangeness PDF than the
DIS experiments. Inclusion of the ATLAS 7 TeV $W/Z$ data leads to
worse $\chi^2_E$ values (higher $S_E$ values) for dimuon SIDIS
production data sensitive to the strangeness PDF. This can be seen in
the comparison of $S_E$ distributions in Fig.~\ref{fig:sn_ct18_ct18z},
where the $S_E$ values for CCFR and NuTeV dimuon data sets are
elevated in the CT18Z fit on the right, as compared to the CT18 fit on
the left, as a consequence of inclusion of the ATLAS $W/Z$ data in the
CT18Z fit.   
(b) The CT18X and CT18Z fits use an $x$-dependent factorization scale
in NNLO DIS cross sections to mimic enhanced higher-order logarithms
at small Bjorken $x$ and small $Q$. This choice results in the
enhanced gluon PDF at small $x$ and reduced gluon at $x \sim 0.01$, as
discussed above. 
Furthermore, the CDHSW data for DIS on heavy nuclei prefer a
somewhat harder gluon PDF at $x >
0.1$  than other data sets.
In the CT18Z fit, we have removed the CDHSW data.
The combination of these choices in the
CT18Z results in the NNLO Higgs production cross section via gluon
fusion that is
reduced by about 1\% compared to the corresponding
CT14 and CT18 predictions. Thus, the various choices made during the
generation of four CT18(A,X,Z) data sets allow us to more faithfully
explore the full range of the PDF behavior at NNLO that is consistent
with the available hadronic data, with implications for electroweak
precision physics measurements and new physics searches at the LHC. 

\begin{acknowledgments}
	The work of J.~Gao was sponsored by the National Natural Science Foundation
	of China under the Grant No. 11875189 and No.11835005. 
	The work at SMU is supported by the U.S. Department of Energy under Grant No. DE-SC0010129. T.~J.~Hobbs acknowledges support from an EIC Center Fellowship. The work of M. Guzzi is supported by the National Science Foundation under Grant No. PHY1820818. The work of C.-P. Yuan was supported by the U.S. National Science Foundation under Grant No. PHY-1719914, and he is also grateful for the support from the Wu-Ki Tung endowed chair in particle physics. 
	
\end{acknowledgments}

\bibliographystyle{JHEP}
%\bibliography{ct18bibtex}

\providecommand{\href}[2]{#2}\begingroup\raggedright\endgroup

\end{document}